\documentstyle[emulateapj]{article}

\begin{document}



\title{Dynamics of Accretion Flows Irradiated by a Quasar}

\author{Daniel Proga\altaffilmark{1}}

\affil{$^1$ Department of Physics, University of Nevada, Las Vegas,
NV 89154, USA, e-mail: dproga@physics.unlv.edu}

\def\LSUN{\rm L_{\odot}}
\def\MSUN{\rm M_{\odot}}
\def\RSUN{\rm R_{\odot}} 
\def\MSUNYR{\rm M_{\odot}\,yr^{-1}}
\def\MSUNS{\rm M_{\odot}\,s^{-1}}
\def\MDOT{\dot{M}}

\newbox\grsign \setbox\grsign=\hbox{$>$} \newdimen\grdimen \grdimen=\ht\grsign
\newbox\simlessbox \newbox\simgreatbox
\setbox\simgreatbox=\hbox{\raise.5ex\hbox{$>$}\llap
     {\lower.5ex\hbox{$\sim$}}}\ht1=\grdimen\dp1=0pt
\setbox\simlessbox=\hbox{\raise.5ex\hbox{$<$}\llap
     {\lower.5ex\hbox{$\sim$}}}\ht2=\grdimen\dp2=0pt
\def\simgreat{\mathrel{\copy\simgreatbox}}
\def\simless{\mathrel{\copy\simlessbox}}

\begin{abstract}
We present the results from axisymmetric time-dependent hydrodynamical
calculations of gas flows which are under the 
influence of gravity of a black hole 
in quasars. We assume that the flows are non-rotating and exposed to
quasar radiation. We take into account X-ray heating and 
the radiation force
due to electron scattering and spectral lines. To compute 
the radiation field, we consider an optically thick, geometrically thin, 
standard accretion disk as a source of UV photons and 
a spherical central object
as a source of X-rays. The gas temperature and ionization state in the flow 
are calculated self-consistently from the photoionization 
and heating rate of the central object.

We find that for a $10^8~\MSUN$ black hole with an accretion luminosity
of 0.6 of the Eddington luminosity the flow settles 
into a steady state and has two components:
(1) an equatorial inflow and
(2) a bipolar inflow/outflow with the outflow leaving the system
along the disk rotational axis. The inflow is a realization
of a Bondi-like accretion flow. The second component
is an example of a non-radial accretion flow which 
becomes an outflow once it is pushed close to the rotational axis 
where thermal expansion and the radiation pressure accelerate it outward.
Our main result is that the existence of
the above two flow components
is robust to the outer boundary conditions and
the geometry and spectral energy distribution of the radiation field.
However, the flow properties are not robust.
In particular, the outflow power and collimation
is higher for the radiation dominated by the UV/disk emission
than for the radiation dominated by the X-ray/central engine emission.
Our most intriguing result is that a very narrow outflow driven by radiation 
pressure on lines can carry more energy and mass than a broad outflow driven by thermal expansion.
\end{abstract}


\section{Introduction}

Active Galactic Nuclei (AGN) in general, but quasars in particular
are very powerful sources of radiation. The AGN luminosity, $L$ is 
typically high compared to its Eddington limit, $L_{\rm Edd}$ 
(0.001~$L_{\rm Edd}\simless L \simless 1~L_{\rm Edd}$).
The spectral energy distribution (SED) of AGN
is very broad. It spans the wavelength range from radio to hard X-rays
and even TeV.
Most of the AGN luminosity is in
the optical--UV--IR bands but some significant fraction is in the X-ray band.

These AGN radiation properties and the AGN central location in 
their host galaxies imply that they play a very important
role in determining the ionization structure and dynamics
of matter not only in their vicinity but also on larger,
galactic and even intergalactic scales
(Ciotti \& Ostriker, 1997, 2001; King 2003; Murray, Quataert, \& Thompson 2005;
Sazonov et al. 2005; Springel, Di Matteo \& Hernquist 2005;
Hopkins et al. 2005; Wang, Chen, \& Hu and references therein). 
There are many indications that support this suggestion, for example
the presence of broad emission lines (BELs) in AGN spectra.

BELs are one of the defining spectral 
features of AGN. They are observed in optical (OPT)
and ultraviolet (UV) spectra and have line wings extending to velocities
up to $10^4~{\rm km~s^{-1}}$. 
It is fairly well established that the primary physical mechanism
for production of BELs is photoionization by the compact
continuum source of AGN. Detailed photoionization calculations yield
relatively tight constraints on the physical conditions
of the emitting gas -- such as the temperature, density, ionization level,
and chemical abundances (e.g., Ferland et al. 1998; Hamman \& Ferland 1999;
Krolik 1999 and references therein).
The width of BELs indicates that the photoionized gas is supersonic.
The shape and position of the line profiles can be explained
by lines forming either in a region
without preferred velocity direction
and with a nearly spherical distribution {\it or} 
at the base of a wind from an accretion disk 
(Murray et al. 1995, hereafter MCGV; Bottorff et. al 1997).
In the latter case, the BELs can form where 
the outflow expansion velocity is low
compared to the rotational velocity (MCGV).

Not all gas in AGN shares the properties of the one responsible for BELs.
Some quasars show broad absorption lines (BALs) which are the most dramatic 
evidence for well-organized outflows in AGN. BALs are almost always 
blueshifted relative to the emission-line rest frame, indicating the presence 
of outflows from the active nucleus, with velocities as large as 
0.2~c (e.g., Turnshek 1998). BALs are observed not only in the UV but also 
in  the X-rays. For example, Chartas, Brandt \& Gallagher (2003) 
discovered a very broad absorption line in the X-ray spectrum of 
PG~1115+80. 
Other evidence for AGN outflows include narrow absorption lines (NALs). 
UV spectra of some quasars show NALs which are blueshifted by as much 
as $\sim$~50000~${\rm km~s^{-1}}$ (Hamann et al. 1997). 
NALs are found much more commonly in the UV spectra of Seyfert galaxies
than of quasars but  in Seyfert galaxies the lines are blueshifted 
only by several 100~${\rm km~s^{-1}}$ (Crenshaw et al. 1999).  
As BALs, NALs are observed not only in
the UV but also in  the X-rays. For example, 
Kaastra et al. (2000) observed NALs due to highly ionized species
in a high-resolution X-ray observation of the Seyfert
galaxy NGC~5548 obtained by Chandra. 
As mentioned above, some argue that 
BELs may also be associated with the base of a disk outflow.

The dynamics of the gas responsible for BALs, BELs, and NALs
can be driven by radiation, even for sub-Eddington sources. 
The driving can be due to radiation pressure or radiation heating, 
or both. The radiation force can overcome gravity for sub-Eddington sources
when the gas opacity is higher than the electron scattering.
The latter is usually used to define $L_{\rm Edd}$. 
The gas opacity can be enhanced by 
scattering of photons by UV spectral lines.
Radiation pressure on spectral lines (line force) 
can be significant, provided gas is moderately ionized and 
can interact with the UV continuum through very many UV line transitions. 
For highly ionized gas, line force is negligible 
because of a lower concentration of ions capable of providing UV line opacity.
In the case of highly or fully ionized gas, an outflow still can be produced
if the gas heating is efficient enough for the thermal energy 
to exceed the gravitational energy.
AGN with their broad SED, are systems where both line driving and
radiation heating, in particular, X-ray heating  
can operate. 

The gas dynamics can be also affected by dust. For example
radiation pressure due to the dust opacity can produce an outflow of
dust and also of gas, as the two can be coupled 
(e.g., Phinney 1989; Pier \& Krolik 1992; 
Emmering, Blandford, \& Shlosman 1992; 
Laor \& Draine 1993; 
K\"onigl \& Kartje 1994; Murray et al. 2005
and references therein). 
Matter in AGN is known to be a mixture of gas and dust on large scales
(e.g., Miller \& Goodrich 1985;
Awaki et al. 1991; Antonucci 1984; Blanco, Ward, \& Wright 1990; Krolik 1999
and references therein) but also on subparsec scales as in the Seyfert galaxy
NGC 1068 (e.g., Wittkowski et al. 2004). However,
details of the matter composition 
and microphysics are very complex and poorly understood. 
Additionally, the presence of dust affects the state and dynamics
of gas and vice verse. In particular, it is nontrivial
to include the dust effects on the gas photoionization state.
Therefore studies that consider both matter components 
rely on simplifying assumptions regarding the dynamics and geometry
(e.g., Murray, Quataert, \& Thompson 2005). There are also
studies that focus only on the gas component and consider dynamics
on relatively smaller radii (e.g., Sazonov et al. 2005).
We follow the latter and focus on the gas dynamics.

It is commonly accepted that AGN are powered by disk accretion of matter
onto a supermassive black hole (BH). A wind driven from
such a disk by line force is the most promising hydrodynamical (HD) scenario 
for AGN outflows, especially high luminsity quasars. 
In this scenario, a wind is launched  from the disk by the
local disk radiation at radii where the disk radiation is mostly in the UV
(Shlosman, Vitello \& Shaviv 1985; MCGV).  Such a wind  is continuous and
has mass loss rate and velocity which are capable of explaining
the blueshifted absorption lines observed in many AGN, if the ionization
state is suitable (e.g.,  MCGV, Proga, Stone \& Kallman 2000, PSK 
hereafter; Proga \& Kallman 2004). This wind scenario has 
the desirable feature that for the wind motive power, 
it relies on radiation which is an observable quantity.
However, not all AGN outflows can be explained by line driving
because of too low luminosity or too high ionization state or both
(e.g., Chelouche \& Netzer 2005; 
and Kraemer et al. 2006). Therefore, other mechanisms such as
thermal and magnetic driving are likely important.

Theoretical models predict that X-ray heating can have profound effects on
the gas dynamics in disks. Since X-rays tend to heat low density gas to 
a temperature $T_{\rm C}\sim 10^7$~K, with which matter in an accretion disk is
expected to either puff up and form a static corona, or to produce
a thermal wind, depending on whether the thermal velocity exceeds the local
escape velocity, $v_{\rm esc}$ (e.g. Begelman, McKee and Shields, 1982; 
Ostriker, McKee, \& Klein 1991; Woods et al. 1996;
Proga \& Kallman 2002).
This paper, and this section in particular, are focused 
on non-magnetic processes. Magnetic driving is briefly discussed in \S 4
(see also Proga 2007).

The insights gained from these studies support the radiation driven disk wind 
model for outflows in quasars and likely in other AGN. 
These successes motivate further exploration of radiation effects on gas 
dynamics in and outside AGN using similar methods. The particular issues 
needed attention include:
(1) Can AGN radiation drive an outflow from anywhere else than
an accretion disk (e.g., from a large scale torus believed to
surround an AGN or from an inflowing gas accreting directly onto a BH
or indirectly through an accretion disk)? 
(2) If so,  what is the power of such an outflow
and can AGN radiation control the rate at which matter
is supplied to the AGN accretion disk. More broadly, 
is the AGN radiation an important element of the so-called AGN feedback? 
(3) What is the origin and the dynamics of matter responsible for 
NALs and NELs?

We report here on results from our first phase of studying
gas dynamics in AGN on sub- and parsec-scales. 
We have simplified several aspects of this complex problem. 
For example, we have not taken into account rotation of the flow and 
the magnetic fields. 
Additionally, we have not taken into account dust as we
assume that all the dust has been destroyed. This assumption
is valid interior to the dust sublimation radius,
where the central object radiation is responsible for the dust
destruction. At large radii, still inside our computational
domain, this assumption is less valid but the dust could be
destroyed there by other processes such as dust sputtering.
Our goal is  to set-up 
simulations as simple as possible to study the effects of 
the X-ray heating (important in the so-called
preheated accretion, see Ostriker et al. 1976) and
radiation pressure on gas which is initially captured by a BH. 

We decided to study non-rotating gas because once this gas is initially 
located in the polar region at large radii, it will likely stay 
in this region regardless of the subsequently radial movement with
respect to the BH and its accretion disk,
i.e, it will not settle down on the accretion disk. 
Therefore non-rotating gas can be exposed to 
the direct disk radiation for a wide range of radii and radial velocities. 
For these reasons, non-rotating gas
can easily absorb, re-emit, and scatter the disk radiation
and be responsible for line emission and absorption in AGN.

We use the methods developed by PSK to study gas dynamics 
on sub-parsec- and parsec-scales in AGNs. 
We consider an axisymmetric HD flow accreting onto a supermassive BH. 
The flow is non-spherical because it is irradiated 
by an accretion disk and spherical corona. The disk radiation flux 
is the highest along the  disk rotational axis and is gradually decreasing 
with increasing polar angle, $\theta$ as $\cos{\theta}$.
The corona radiation is isotropic.
We take into account the radiation heating and cooling, radiation pressure due
to the electron scattering and spectral lines.
We adopt a simplified treatment of photoionization,
and radiative cooling and heating that allow us to compute 
self-consistently the ionization state, and therefore the line force, 
in the flow.

We concentrate on assessing what fraction of the inflow from large
radii reaches the vicinity of a BH and what fraction is turned into
an outflow. We also assess the impact of thermal expansion and
radiation pressure on driving the outflow and on 
the outflow properties such as the two-dimensional structure, 
thermal and kinetic power and temporal behavior.
We describe our calculation in Section 2. We present our results in Section 3.
The paper ends, in section 4, with discussion and our conclusions.

\section{Method}

\subsection{Hydrodynamics}

To compute the structure and evolution of a flow irradiated by
an AGN, we solve the equations of hydrodynamics
\begin{equation}
   \frac{D\rho}{Dt} + \rho \nabla \cdot {\bf v} = 0,
\end{equation}
\begin{equation}
   \rho \frac{D{\bf v}}{Dt} = - \nabla P + \rho {\bf g}
 + \rho {\bf F}^{\rm rad}
\end{equation}
\begin{equation}
   \rho \frac{D}{Dt}\left(\frac{e}{ \rho}\right) = -p \nabla \cdot {\bf v}
   + \rho \cal{L}  ,
\end{equation}
where $\rho$ is the mass density, $P$ is the gas pressure,
${\bf v}$ is the velocity, $e$ is the internal energy density,
$\cal{L}$ is the net cooling rate,
${\bf g}$ is the gravitational acceleration of the central object,
${\bf F}^{\rm rad}$ is the total radiation force per unit mass.
We adopt an adiabatic equation of state
$P~=~(\gamma-1)e$, and consider models with the adiabatic index, $\gamma=5/3$.

To solve eqs. 1-3, we use the ZEUS-2D code (Stone \& Norman 1992)
extended by PSK. 
We perform our calculations in spherical polar coordinates
$(r,\theta,\phi)$ assuming axial symmetry about the rotational axis
of the accretion disk ($\theta=0^o$). 

Our standard computational domain is defined to occupy 
the angular range $0^o \leq \theta \leq 90^o$ and the radial range
$r_{\rm i}~=~500~r_\ast \leq r \leq \ r_{\rm o}~=~ 2.5~\times~10^5~r_\ast$, 
where $r_\ast=3 r_{\rm S}$ is the inner radius of the disk
around a Schwarzschild BH with a mass, $M_{\rm BH}$ and radius 
$r_{\rm S}=2GM_{\rm BH}/c^2$.
The $r-\theta$ domain is discretized into zones.
For $r_{\rm o}= 2.5~\times~10^5~r_\ast$ and  $1~\times~10^6~r_\ast$, 
our numerical resolution in the $r$ direction consists of 140 and 180
zones, respectively.
We fix the zone size ratio, 
$dr_{k+1}/dr_{k}=1.04$ (i.e., the zone spacing is increasing outward). 
Gridding in this manner ensures good spatial resolution close to
the inner boundary at $r_{\rm i}$.
In the $\theta$ direction, our standard numerical resolution consists
of 50 but we also performed simulations with 100 zones.
Zone size ratio is always $d\theta_{l}/d\theta_{l+1} =1.0$ (i.e.,
grid points are equally spaced).

For the initial condition, we typically assume spherical symmetry
set all HD variables to constant values everywhere in 
the computational domain, i.e., $\rho(r,\theta)=\rho_{\rm 0}$
$e(r,\theta)=e_{\rm 0}$, 
$v_r(r,\theta)=v_{r,0}=0$, 
$v_\theta(r,\theta)=v_{\theta,0}=0$, and
$v_\phi(r,\theta)=v_{\phi,0}=0$.
For comparison, we performed simulations 
where we initially set HD variables
to values which resulted from one-dimensional simulations
without radiation pressure. 

We specify the boundary conditions in the following way. 
At the pole (i.e., $\theta = 0^\circ$), we apply an axis-of-symmetry boundary 
condition while for $\theta=90^\circ$, we apply reflecting boundary
conditions.
For the inner and outer radial boundaries, 
we apply an outflow boundary condition (i.e., 
to extrapolate the flow beyond the boundary, we set values
of variables in the ghost zones equal to the values in the corresponding
active zones, see Stone \& Norman 1992 for more details). 
To represent steady conditions at the outer radial boundary, 
during the evolution of each model we continue to apply the constraints that 
in the last zone in the radial direction, 
$v_\theta(r_{\rm o},\theta) =v_\phi(r_{\rm o},\theta) =0$, 
and the density and internal energy are fixed at constant values 
at all times.  
We allow $v_r$ to float. 
We have found that this technique produces a solution that relaxes 
to a steady state solution for both  spherically symmetric accretion
and non spherical accretion with an outflow (see also Proga \& Begelman 2003a).
Our approach then is to mimic the situation where there is
always gas available for accretion.

\subsection{The radiation field and force}

The geometry and assumptions needed to compute the radiation
field from the disk and central object are similar to those
considered by PSK but here we make some simplifications to the earlier
model.
The disk is flat, Keplerian, geometrically-thin and optically-thick.
The disk photosphere is  coincident with the $\theta=90^o$ axis.
We specify the radiation field of the disk  by assuming that the temperature
follows the radial profile of the optically thick accretion disk
(Shakura \& Sunyaev 1973), and therefore depends on
the mass accretion rate in the disk, $\dot{M}_{\rm D}$, the mass of the BH,
$M_{\rm BH}$  and  the inner edge of the disk, $r_\ast$, 
In particular, the total accretion luminosity, 
$L=2 \eta G M_{\rm BH} \MDOT_{\rm a}/r_{\rm S}$,
where $\eta$ is the rest mass conversion efficiency.

The geometry of the central engine in AGNs is poorly known. 
As in PSK, we consider the central engine as the most inner part of 
the accretion disk plus an extended corona.
We refer to the corona as the central object.
The inner radius of the computational domain is large compared to 
the radius of the central object.
Therefore the central object can be approximated as a point source 
located at $r=0$.
We express the disk luminosity, $L_{\rm D}$ and  
the central object luminosity, $L_\ast$ in units of 
the total accretion luminosity, i.e.,  $L_{\rm D}=f_{\rm D} L$ and
$L_\ast=f_\ast L=(1-f_{\rm D}) L$.
For simplicity, we assume that the disk emitts only in the UV, whereas
the central object emitts only the X-rays. i.e.,
the system UV luminosity, $L_{\rm UV}=f_{\rm UV} L=L_{\rm D}$ 
and
the system X-ray luminosity, $L_{\rm X}=f_{\rm X} L=L_\ast$
(in other words $f_{\rm UV}=f_{\rm D}$ and $f_{\rm X}=f_\ast$).

With the above simplifications, only the central object
radiation is responsible for ionizing the flow to a very high state.
The central radiation contributes to the radiation force 
due to electron scattering but does not contribute to line driving.
On the other hand, the disk radiation contribute
to the radiation force due to both electrons and lines.
Note that we assume that
the luminosity in the remaining bands, mainly optical and infrared,
is the part of the luminosity that 
does not change the dynamics of the wind and set it  to zero.

To  evaluate, line force, we generally follow PSK who 
used a modified Castor, Abbott \& Klein method (Castor, Abbott, \& Klein
1975;
see also Proga, Stone, Drew 1998, 1999).  
The line force at a point defined by the position vector $\bf r$ is
\begin{equation}
{\bf F}^{\rm rad,l}~({\bf{r}})=~\oint_{\Omega} M(t)
\left(\hat{n} \frac{\sigma_e I({\bf r},\hat{n}) d\Omega}{c} \right)
\end{equation}
where $I$ is the frequency-integrated continuum intensity in the direction
defined by the unit vector $\hat{n}$, and $\Omega$ is the solid angle
subtended by the disk and central object at the point.
The term in brackets is the electron-scattering radiation force,
$\sigma_e$ is  the mass-scattering coefficient for free electrons,
and $M(t)$ is the force multiplier (Castor et al. 1975) 
-- the numerical factor which
parametrizes by how much spectral lines increase the scattering
coefficient. In the Sobolev approximation, $M(t)$ is a function
of the optical depth parameter
\begin{equation}
t~=~\frac{\sigma_e \rho v_{\rm th}}
{ \left| dv_l/dl \right|},
\end{equation}
where $v_{\rm th}$ is the thermal velocity,
and $\frac{dv_l}{dl}$ is the velocity gradient along the line of sight,
$\hat{n}$. 

Evaluation of the radiation force due to an non-isothermal extended
source such as an accretion disk is relatively expensive (see PSK, 
Proga et al. 1999).
However, we consider here the flow relatively
far from both the central object and the most luminous part
of the disk. Therefore, we simplify the
force evaluation as follows. First, we consider only the radial
component of the force, ${\bf F}^{\rm rad,l}_r$
and set the other components to zero.
Second, we compute the velocity gradient by taking
into account only the dominant term:
gradient of the radial velocity
along the radial direction, i.e. $\frac{dv_r}{dr}$.

With these simplifications and an assumption that the flow does not change 
the geometry of the radiation field, we can evaluate the radiation force 
in two steps: 
(1) calculation of the velocity gradient in the $\hat{n}_r$ direction and 
then the optical depth parameter $t$; (2) calculation of the parameters of the
force multiplier using a current value of the photoionization
parameter, $\xi$ adopting results of Stevens \& Kallman (1990).
Then we calculate the radiation force exerted by radiation along 
the radial direction. Our approach to  evaluate the radiation forces 
is simplified compared to the approach used in PSK  because
we do not integrate the radiation force over the solid
angle subtended by the radiant surface but instead use 
an analytic approximation (compare eqs. 4 and 8), 
and we do not correct the radiation force for the optical depth effects.

Below we describe in more detail, our calculations of the force multiplier 
and the radiation force for various conditions in the flow. For $r \gg r_\ast$, 
in the optically thin case, the radial radiation flux from a disk and central 
object can be written as
\begin{equation}
{\cal F}_{\rm D}(r,\theta)   = 2 \cos{\theta} \frac{L_{\rm D}}{4\pi r^2},
\end{equation}
and
\begin{equation}
{\cal F}_\ast(r)~=~\frac{L_\ast}{4\pi r^2},
\end{equation}
respectively. Eq. 6 approximates the numerical evaluation
of the radiation flux integral over the solid
angle subtended by the disk surface 
(see eqs B2 and B3 in Proga, Stone \& Drew 1998) while
eq. 7 is the exact solution for the radiation flux
for a point source.
With eqs. 6 and 7, one can show that eq. 4 reduces to
\begin{equation}
{\bf F}_r^{\rm rad}~(r,\theta)= 
\frac{\sigma_e L}{4\pi r^2 c}\left[f_\ast+2\cos{\theta}f_{\rm D} (1+M(t))\right]
\end{equation}

To compute the force multiplier, we adopt Castor et al.'s (1975) 
analytical formula
modified by Owocki, Castor \& Rybicki (1988 see also Proga et al. 1998)
\begin{equation}
M(t)~=~k t^{-\alpha}~
\left[ \frac{(1+\tau_{\rm max})^{(1-\alpha)}-1} {\tau_{\rm max}^{(1-\alpha)}} \right]
\end{equation}
where $k$ is proportional to the total number of lines, $\alpha$ is the
ratio of optically-thick to optically-thin lines,
$\tau_{\rm max}=t~\eta_{\rm max}$ and $\eta_{\rm max}$ is a parameter determining
the maximum value, $M_{\rm max}$ achieved for the force multiplier.
Equation 9 shows the following limiting behavior:
\begin{eqnarray}
\lim_{\tau_{\rm max} \rightarrow \infty}~M(t) & = & k t^{-\alpha} \\
\lim_{\tau_{\rm max} \rightarrow 0}~M(t) & = & M_{\rm max},
\end{eqnarray}
where $M_{\rm max} = k (1-\alpha)\eta_{\rm max}^\alpha$. 

To compute, the parameters of $M(t)$ taking into account the effects of 
ionization, we follow PSK. In particular, we compute the parameters of the
line force using a value of the photoionization parameter, $\xi$ and 
the analytical formulae for $k$ and $\eta_{\rm max}$ 
from Stevens \& Kallman (1990): 
\begin{equation}
k= 0.03 + 0.385 \exp(-1.4~\xi^{0.6}),
\end{equation}
and
\begin{equation}
\log_{10} \eta_{\rm max} = \left\{ \begin{array}{ll}
6.9~\exp(0.16~\xi^{0.4})
& {\rm for}~\,~
\log_{10} \xi~\leq~0.5 \\
 & \\
9.1 \exp(-7.96~\times10^{-3}\xi)
& {\rm for} ~\,~
\log_{10}~\xi~>~0.5  \\
\end{array}
\right.
\end{equation}
The  parameter $\alpha=0.6$ and does not change with $\xi$.

Our procedure to calculate $M(t)$ is computationally efficient and gives 
approximate estimates 
for the number and opacity distribution of spectral lines for a given $\xi$ 
without detail information about the ionization state.
However, it is just an estimate because
Stevens \& Kallman (1990) results were computed for the case
where a 25 000~K stellar atmosphere was irradiated by
a 10 keV bremsstrahlung ionizing spectrum whereas we consider
gas  not of a stellar atmosphere and its
temperature, without irradiation,  depends of the outer boundary conditions.
Additionally, the AGN spectrum is different
from a bremsstrahlung spectrum. As we describe below, 
the parameters $k$ depends on the gas temperature (i.e., eq. 17).
Additionally 
this and other parameters of $M(t)$ depend also on the SED of the ionizing
and UV radiation. In particular, $\log_{10} \xi= -0.5$
and the UV emission caused by the low energy extension of 
the 10 keV bremsstrahlung,
$M_{\rm max}\approx 300$ whereas for an AGN-like UV spectrum, i.e.,
a power law with the energy index equal to 1, 
$M_{\rm max}\approx 8000$ 
(T. Kallman private communication). For comparison, 
eqs. 12 and 13
yield $M_{\rm max}\approx 3400$ also for $\log_{10} \xi= -0.5$.
A detail investigation of these and other dependencies
is beyond the scope of this study but it is important
to carry out such a study in the future. 

To calculate the photoionization parameter needed in eqs. 12 and 13 
and in evaluation of the net radiative cooling, we 
use the formula:
\begin{equation}
\xi = \frac{4 \pi {\cal F}_{\rm X}}{n},
\end{equation}
where ${\cal F}_{\rm X}$ is the local  X-ray flux, $n$
is the number density of the gas ($={\rho}/({m_p \mu })$ , where
$m_p$ is the proton mass, and $\mu$ is the mean molecular weight).
We consider models with $\mu=1$.
The  local X-ray flux is corrected for
the optical depth effects:
\begin{equation}
{\cal F}_{\rm X}={\cal F}_\ast \exp(-\tau_{\rm X}),
\end{equation}
where $\tau_{\rm X}$ is the X-ray optical depth.
We estimate $\tau_{\rm X}$ between the central
source and a point in a flow from:
\begin{equation}
 \tau_{\rm X}= \int_0^{r} \kappa_{\rm X} \rho~dr,
\end{equation}
where $\kappa_{\rm X}$ is the absorption coefficient,
and $r$ is the distance from the central source.

The flow physical conditions and consequently the line force depend
not only on $\xi$. In particular, the gas temperature
and ionization state can be affected by adiabatic heating or cooling. 
For example, the line force can be neglected
even in the region of low $\xi$ if this region is hot due 
to adiabatic compression. Therefore, we first compute
the line force parameters based on $\xi$ and then 
correct the parameters for the gas temperature effects.

Specifically, we compute the parameter $k$ using the expression
\begin{equation}
\log_{10}{k} = \left\{ \begin{array}{ll}
                   -0.383  & {\rm for} \,     \log_{10} T \leq 4 \\
-0.630 \log_{10}{T}+2.138  & {\rm for} \, 4 < \log_{10} T \leq4.75 \\
-3.870 \log_{10}{T}+17.528 & {\rm for} \,     \log_{10} T > 4.75 \\
\end{array}
\right.
\end{equation}
We adopt this value of $k$ if it is smaller than the one
obtained by using expression $12$. Expression $17$ is based on 
detailed photoionization calculations performed using  the XSTAR code
(T. Kallman private communication). 

These expressions for the parameters of the force multiplier
predict  that $M_{\rm max}$ increases gradually from $\sim 2000$
to $5000$ as $\xi$ increases from 0 to $\sim 3$ and then drops to $\sim 0.1$
at $\xi=1000$. The line force becomes  negligible
for $\xi \simgreat 100$ because then $M_{\rm max} \sim 1$.
The line force becomes also negligible
if $T>10^5$~K for any $\xi$.

\subsection{Radiation heating and cooling}

To calculate the gas temperature we also follow PSK, who 
assumed that the gas is optically thin to its own cooling radiation. 
The net cooling rate depends
on the  density, $\rho$, the temperature, $T$, the ionization parameter
$\xi$, and the characteristic temperature of the X-ray radiation $T_{\rm X}$.
We refer a reader to PSK and Proga \& Kallman (2004) for more details.

We calculate the evolution of the internal energy and therefore
we specify the initial and outer boundary conditions for $e$. 
We do this in the following way: we specify  the gas temperature at the outer 
boundary (in the last grid zone in the radial direction), $T_{\rm 0}$,
by setting it to the Compton temperature, $T_{\rm C}=0.25~T_{\rm X}$.
Then we calculate the internal energy from
\begin{equation}
e_{\rm 0}=\frac{\rho_{\rm o} k T_{\rm 0}}{\mu m_p (\gamma -1)}.
\end{equation}

\section{Results}

We assume the mass of the non-rotating BH,
$M_{\rm BH}~=~10^8~\rm \MSUN$ and
the disk inner radius, $r_\ast=~3~r_{\rm S}~=~8.8~\times~10^{13}$~cm throughout
this paper.
We consider the rest mass conversion efficiency $\eta=~0.0833$.
We set $\MDOT_{\rm a}$ to be 
$10^{26}~{\rm g~s^{-1}}$ (=1.6~$\MSUNYR$).
These system parameters yield 
the accretion luminosity, 
$L = 7.5\times10^{45}~{\rm erg~s^{-1}}(=2\times10^{12}~\LSUN$). 
This luminosity corresponds to the Eddington number, $\Gamma=0.6$.
We define the Eddington number as $L$
in units of the
Eddington luminosity for the Schwarzschild BH,
$L_{\rm Edd}= 4 \pi c G M_{\rm BH}/\sigma_e$ [i.e.,
$\Gamma\equiv L/L_{\rm Edd}= (\sigma_e \MDOT_{\rm a})/(8\pi c r_{\rm S}$)].
To determine the radiation field, we specify 
also $f_{\rm UV}$ and $f_{\rm X}$. We note that we fix 
$\MDOT_{\rm a}$ during our simulations, i.e. $\MDOT_{\rm a}$ does not depend on
the mass flux through the inner boundary, 
$\MDOT_{\rm in}(r_{\rm i})$ 
(see eq. 20 for the formal definition of $\MDOT_{\rm in}(r_{\rm i})$
and \S 4 for discussion of this aspect of the calculations).

The parameters $f_{\rm UV}$  and $f_{\rm X}$ (and also $f_{\rm D}$ and $f_\ast$)
specifying the radiation field
are our free parameters which we focus on exploring the most in this paper.
We consider three cases: case A with $f_{\rm UV}=0.5$  and $f_{\rm X}=0.5$, 
case B with $f_{\rm UV}=0.8$  and $f_{\rm X}=0.2$, and
case C with $f_{\rm UV}=0.95$  and $f_{\rm X}=0.05$
(see Table 1 for summary of our runs). 
The spectral energy distribution of the ionizing radiation is not
well known, our choice of values for $f_{\rm UV}$ and $f_{\rm X}$
is guided by the observational results from  Zheng et al. (1997) 
and Laor et al. (1997).

To calculate the gas temperature, we assume the temperature of
the X-ray radiation, $T_{\rm X}=8\times10^7$~K 
(e.g., Sazonov, Ostriker \& Sunyaev 2005 and references therein)
and the line cooling parameter
$\delta=1$ (see Blondin  1994 and PSK).
The force multiplier depends only formally on the thermal speed,
$v_{\rm th}$. Therefore to compute $t$, 
we set to 20 ${\rm km~s^{-1}}$, i.e., the thermal speed of
a hydrogen atom at the temperature of 25000~K for which
the parameters of the force multiplier were computed
(Stevens \& Kallman 1990). Finally,
the attenuation of the X-rays is calculated using
$\kappa_{\rm X}= 0.4~{\rm g^{-1}~cm^2}$ for all $\xi$.
The resulting optical depth corresponds to the Thomson optical depth.

For our parameters, both the so-called Compton radius,
$R_{\rm C}\equiv G M_{\rm BH} \mu m_p/k T_{\rm C}$ = 8$\times10^{18}$~cm 
= $9\times10^4~r_\ast$ 
and the Bondi radius, $R_{\rm B}=G M_{\rm BH}/c_\infty^2= 4.8 \times 10^{18}$~cm = 
$5.5 \times 10^4~r_\ast$ (Bondi 1952) are inside our computational domain. We assumed here
that the gas temperature  at infinity $T_\infty=T_{\rm C}=2\times10^7$~K so that
the sound speed at infinity, 
$c_\infty=(\gamma k T_{\rm C}/\mu m_p)^{1/2}$= 4$\times10^7$~cm~$\rm s^{-1}$
and $R_{\rm C}=\gamma R_{\rm B}$. For the isothermal flow, the Bondi
accretion rate is 
$\MDOT_{\rm B}=3.3\times10^{25}{\rm g~s^{-1}}$= 0.52~$\MSUNYR$.
We note that the dust sublimation radius, $R_{\rm sub}$=~2.6$\times10^{18}$~cm~
=3$\times10^4~r_\ast$ (e.g., Murray et al. 2005, eq. 41) 
is also inside the domain.

We consider the situation where 
the BH gravity is reduced by the $\theta$-dependent radiation force.
Therefore, the effective gravity is also $\theta$ dependent.
Using eqs. 6 and 7, the location where thermal energy equals 
the effect gravitational energy (i.e., a modified Compton radius) 
can be written as
\begin{equation}
{\bar R}_{\rm C} \equiv R_{\rm C} [1-\Gamma (2 \cos{\theta} f_{\rm D}+f_\ast)].
\end{equation}

Figure~1 presents the results for cases A, B, and C
with $\rho_{\rm 0}=10^{-21}~\rm g~cm^{-3}$, 
$v_{r, {\rm 0}}=v_{\theta, {\rm 0}}=v_{\phi, {\rm 0}}$=0.
At the outer radial boundary,
during the evolution of each model we continue to apply the constraints that
the density and internal energy (the latter is computed for
$T_{\rm 0}=T_{\rm C}$) are fixed at constant values
at all times. 
The figure shows the instantaneous density and temperature
distributions, and the poloidal velocity field of the models.
Additionally, it shows ${\bar R}_{\rm C}$ and the contours of the Mach number , 
$M~\equiv~\sqrt{v^2_r+v^2_\theta}/c_s$,
where $c_s=\gamma P/\rho$ is the sound speed.
For all three cases,
the flow settles quickly into a steady state 
(within $\sim$ a few $\times 10^{12}$~s
which correspond to a few dynamical time scales
at the outer radius, $\tau=(r_o^3/G M_{\rm BH})^{1/2}=9 \times 10^{11}$~s ). 
The steady state consists of two flow components
(1) an equatorial inflow and
(2) a bipolar inflow/outflow with the outflow leaving the system
along the pole. The outflow is collimated by the infalling gas.
Our calculations capture the subsonic and supersonic
parts of both the inflow and outflow.
Although the same components can be identified in the three cases
their size, density and temperature, and the degree  of outflow
collimation depend on the SED and the geometry of the radiation field.

In run A, X-ray heating is the strongest 
and it accelerates an outflow to the maximum velocity of 700~km~$\rm s^{-1}$.
Although, the gas temperature decreases in the inflow down to some
radius, it is not low enough for line force to be significant.
The outflow collimation by the infall increases with increasing
radius. This tendency is clearly seen in the other two cases.

For run B, the outflow is strongly pushed toward the pole 
at $r\approx 1.2\times10^5~r_\ast$ and the flow pattern
resembles a bottle neck where the density increases by a couple
of orders of magnitude. The gas temperature drops there to the level
below $10^5$~K. In this low temperature region line driving
accelerates  an outflow to the velocity up to 4000~km~$\rm s^{-1}$ 
which is significantly higher than
the velocity of the thermal outflow in run~A. 

The outflow collimation is very strong for run C where X-ray heating
is the smallest. The gas is siphoned off within a very narrow
channel along the pole. In this case, the fraction of the computational
domain occupied by the inflow is the largest.
The shape and size of the sonic surface reflects the fact
that runs A and B are similar to each other, 
but they both differ significantly
from  run C.

To quantify the properties of the inflow and inflow/outflow components,
we computed three radial mass rates as a function of radius:
(1) the net rate
\begin{equation}
\MDOT_{\rm net}(r) =
4 \pi r^2 \int_{0^\circ}^{90^\circ} \rho v_r \sin \theta d\theta,
\end{equation}
(2) the inflow rate 
\begin{equation}
\MDOT_{\rm in}(r) =
4 \pi r^2 \int_{0^\circ}^{90^\circ} \rho v_r \sin \theta d\theta~~~~{\rm for}~v_r~<~0,
\end{equation}
and (3) the outflow rate
\begin{equation}
\MDOT_{\rm out}(r) =
4 \pi r^2 \int_{0^o}^{90^\circ} \rho v_r \sin \theta d\theta~~~~{\rm for}~v_r~>~0.
\end{equation}
We also computed the outflow power carried out in the form
of the kinetic energy
\begin{equation}
{P}_{\rm k}(r) =
2 \pi r^2 \int_{0^o}^{90^\circ} \rho v^3_r \sin \theta d\theta~~~~{\rm for}~v_r~>~0
\end{equation}
and in the form of the thermal energy
\begin{equation}
{P}_{\rm th}(r) =
4 \pi r^2 \int_{0^o}^{90^\circ} e v_r \sin \theta d\theta~~~~{\rm for}~v_r~>~0.
\end{equation}
These quantities are shown in 
Figs. 2, 3, and 4 for run A, B, and C, respectively.

Fig. 2 shows that
the outflow in case A is indeed thermal:  $P_{\rm k}<P_{\rm th}$ for all radii;
the outflow starts at radii large compared to $R_{\rm C}$, 
its properties are strong functions of radius, i.e.,
$\MDOT_{\rm out}$, $P_{\rm k}$, and $P_{\rm th}$ increase with radius.
In particular, almost all the gas inside $r<2 \times 10^4~r_\ast$ 
ends up leaving the domain through the inner boundary as there
is no  or a very weak outflow driven from small radii.

The reduction of $f_\ast(=f_{\rm X})$ compared to $f_{\rm D}(=f_{\rm UV})$
has a few significant consequences on the outflow properties:
in cases B and C the outflow rate is appreciable even at small
radii (see figs. 3 and 4). 
In particular, in case C the net rate at large radii is almost
an order of magnitude  smaller than the inflow rate. The latter is almost
canceled out by the outflow rate. The overall power of the outflow
is $\simgreat 2$ orders of magnitude higher in cases B and C
than in case A. The relative weakness of thermal driving
for lower $f_{\rm X}$ cases is obvious when $P_{\rm th}$ is compared
to $P_{\rm k}$. In run B, $P_{\rm th}$ increases with increasing radius for 
$r\simless 10^5$.
But even at its maximum, $P_{\rm th}$ is 3 orders of magnitude
lower than $P_{\rm k}$. In run C, the kinetic dominance
is even greater: 
$P_{\rm k}$ is higher than $P_{\rm th}$ by 4 orders of magnitude.

The dependence of our results on the SED ($f_{\rm X}$ and $f_{\rm UV}$) 
and the radiation geometry ($f_\ast$ and $f_{\rm D}$) can be explained 
by the fact
that as $f_{\rm X}$ and $f_\ast$ decrease the X-ray heating also decreases
whereas the radiation flux in the polar region increases
compared to the flux at high $\theta$ (see eq. 8). Consequently, 
an increasing fraction of the inflow from large radii is pushed toward 
the polar region where radiation pressure exceeds gravity. 
This compressed inflow is siphoned off by radiation pressure.
Perhaps the most intriguing result is that the large scale
flow pattern can be very misleading if one wants
of assess the relative strength of inflow and flow.
Visual inspection of Fig. 1 could
lead to the conclusion that the outflow in run C is much
less significant that the outflow in run A because in run C
the outflow is very narrow and barely seen. However, the 
opposite is true as shown in Table 1. The increase in the mass
outflow rate with decreasing $f_{\rm X}$ is due to increasing
$f_{\rm UV}$ and the increasing efficiency of radiative driving
at $f_{\rm X}$.
 
Table~1 shows that the inflow rate at $r_{\rm o}$ increases with 
decreasing $f_{\rm X}$.
This trend is consistent  with accretion theory,  in particular with
Bondi's estimate for the accretion rate, 
$\MDOT_{\rm B} \propto T_\infty^{-3/2}$:   
weaker X-ray heating results in an increase of the accretion rate because the gravitational sphere of 
influence over gas pressure is larger.
However, a higher $\MDOT_{\rm min}$ at large radii does not mean that the inflow
rate at small radii must be higher too. 
We find that the mass outflow rate also increases with decreasing $f_{\rm X}$. 
Additionally, the efficiency of turning
an inflow into an outflow is also sensitive to $f_{\rm X}$:
$\left|\MDOT_{\rm out}(r_{\rm o})/\MDOT_{\rm in}(r_{\rm o})\right|=$ 3/4, 5/8, and 8/9, for $f_{\rm X}$= 0.5, 0.2, and 0.05, respectively.
As we mentioned above, the decrease from high to low values of $f_{\rm X}$ 
corresponds to a transition from a thermally to radiation driven outflow.
We conclude then that thermal driving is less efficient than radiation driving.

One of the parameters specifying our problem is
the gas temperature at the outer boundary, $T_{\rm 0}$. 
For the runs A, B, and C, we set the value of $T_{\rm 0}$ equal
to the Compton temperature. This value of the gas temperature is  expected if the gas is optically
thin and heating by X-rays from the central object. However, it is possible
that the gas at large radii is heated also by other processes (e.g., shocks)
or the X-ray heated is reduced by optical depth effects. Therefore,
$T_{\rm 0}$  
can be decoupled from $T_X$ and be lower or higher than $T_{rm C}$.
To check how are results depend on the assumed value of $T_{\rm 0}$,
we rerun model B with $T_{\rm 0}$ set to 1/10, 1/3, and 3 of $T_{\rm C}$,
runs B1, B2, and B3, respectively.

As shown in Table~1, $\MDOT_{\rm in}(r_{\rm o})$ 
and $\MDOT_{\rm net}(r_{\rm i})$ increase with increasing $T_{\rm 0}$. 
However, one would expect the opposite trend because
accretion theory predicts that $\MDOT_{\rm B} \propto T_\infty^{-3/2}$
($T_{\rm 0}$ is a proxy of $T_\infty$ in the simulations).
This discrepancy is due to the fact that the above scaling law is derived
for the Bondi accretion problem where the gas is adiabatic or isothermal,
i.e., where the gas temperature at the critical point increases with 
$T_\infty$.
However, in our problem, the gas is not adiabatic as we
allow for radiative cooling and heating. This has an important
effect on the gas properties (e.g., Zel'dovich \& Novikov 1971) . 
In particular, for 
$T_{\rm 0}> T_{\rm C}$, radiative processes lead to cooling
as the gas accretes. Consequently, the sonic radius is smaller
than that for $T_{\rm 0}=T_{\rm C}$. 
Similarly, for $T_{\rm 0}<T_{\rm C}$, radiative processes lead to heating
as the gas accretes. Consequently, the sonic radius is larger
than that for $T_{\rm 0}=T_{\rm C}$.
Therefore, our results for various $T_{\rm 0}$ reflect
the basic physical property of accretion flows:
flows with a higher temperature at the sonic radius have lower
accretion rates than flows with a lower temperature
because the gravitational sphere of 
influence over gas pressure grows with decreasing 
temperature at the sonic radius.

\section{Discussion and Conclusions}

We have calculated a series of models for non-rotating flows 
which are under the influence of super massive BH gravity and
radiation from an accretion disk surrounding
the BH. Our numerical approach allows for the self-consistent
determination of whether the flow is gravitationally captured by the BH
or driven away by thermal expansion or radiation pressure.
We find that the flow settles quickly 
into a steady state and has two components
(1) an equatorial inflow and
(2) a bipolar inflow/outflow with the outflow leaving the system
along the pole. The first components is a realization
of Bondi-like accretion flow. The second component
is an example of a non-radial accretion flow which 
becomes an outflow once it is pushed close to the rotational axis of the disk
where thermal expansion and radiation pressure can accelerate 
the flow outward.
Our main result is that the existence of
the above two flow components
is robust 
yet their properties are
sensitive to the geometry and SED of the the radiation field
and the outer boundaries.
In particular, the outflow power and the degree of collimation
is higher for the model with the radiation dominated by the UV/disk emission
(run C)
than for model with the radiation dominated by 
the X-ray/central engine emission (run A).
This sensitivity is related to the fact that thermal
expansion compared to radiation pressure
drives a weaker, less-collimated outflow.

Our results will likely change if we allow for significant gas rotation. 
In particular, 
the gas will tend to converge toward the equator due to the combination of 
the centrifugal and gravitational forces. This, in turn, will likely weaken 
the outflow in the polar region because less gas will be pushed toward 
the polar region. Adding rotation is indeed
an important next step. However, within our framework,
it requires the introduction of additional free parameters.
In particular, we would need to prescribe the rotational
rate at the outer boundary. I plan to report on our results from
some simulations with rotation in a companion paper.

The treatment of the inner and outer boundaries can affect  the results
of these simulations. We set up the outer boundary in the spirit
of representing steady conditions. However, in real systems
but also in our simulations, the conditions at large radii
are complex.
In particular, we fix the density and internal
energy at $r_{\rm o}$ at all times. This produces self-consistent
results for accretion flows but not for outflows
because the outflow density and internal energy at $r_{\rm o}$
do not have to be as those we set there through implementation
of the outer boundary conditions. Reruns of our simulations on a larger 
computational domain and with the outer boundary conditions as predicted
by simulations of galaxy evolution such as those performed by 
Springel, Di Matteo \& Hernquist (2005) would be very important.

It is also important to explore the effects of dust on our results.
As we mentioned in \S3, the dust sublimation radius is inside
our computational domain. The presence of the dust at large radii
can reduce the matter density there, because the radiation pressure
on dust can accelerate the matter outward. Consequently, the mass
supply rate and density at small radii can decrease.
This in turn can lead to higher gas temperature at small radii. 
If the dust reduces the gas temperature at large radii,
based on our simulation with various $T_{\rm 0}$, we expect
similar changes.
In summary,
we expect that inclusion of dust will lead to a lower
net mass accretion rate and higher temperature at small radii
compared to the results presented here.

Our treatment of the inner boundary conditions is one of the simplest
possible: free outflow -- no matter enters the computational domain
through the inner boundary. However, as we mentioned in \S 1, quasars
have winds which are likely driven from relatively small radii.
Therefore, one would need to consider matter entering our
computational domain through the inner boundary. Even if the small
scale quasar winds do not flow very far from the center,
their presence could affect our results. 
We allow the radiation field from the disk and central object
to be attenuated by the gas inside the computational domain.
However, PSK showed that disk winds in AGN can be optically thick
and they can change the geometry and SED of the disk and 
central object radiation at large radii, i.e., 
interior to the inner radius of our computational domain. 

We have performed our simulations in the HD limit and have not
accounted for the effects of magnetic fields. The dynamics of
accretion flows can be significantly affected for magnetic fields.
In particular, outflows can be magnetically driven from
accretion disks (e.g., Blandford \& Payne 1982; Pelletier \& Pudritz 1992
K$\ddot{\rm o}$nigl 1993; 
Emmering, Blandford \& Shlosman 1992;
Contopoulos \& Lovelace 1994; K\"onigl \& Kartje 1994;
Contopoulos 1995;
de Kool \& Begelman 1995;
Bottorff et al. 1997;  Bottorff, Korista \& Shlosman 2000;
Everett, K\"onigl, \& Arav 2002; Proga 2003; Everett 2005).
In fact, most magnetohydrodynamic 
simulations of accretion flows show outflows
(e.g., Uchida \& Shibata 1985; Stone \& Norman 1994;
Ouyed \& Pudritz 1997; Hawley \& Balbus 2002; Proga \& Begelman 2003b;
De Villier et al. 2004; McKinney \& Gammie 2004, and references therein).
Additionally, an outflow can be produced even 
from a flow with angular momentum so low
that, if not for the effects of magnetic fields, 
it would accrete directly onto a BH without forming a disk (Proga 2005).
Thus, magnetic fields can directly and indirectly change our results,
e.g., magnetic disk winds, similarly  to radiation driven disk winds,
can change the SED and they can interact with other flow components. 
We expect that flow interaction can lead to time-dependent evolution 
and formation of inhomogeneous inflows and outflows (e.g., Proga \& Begelman
2003b). Therefore, it would be a natural step to repeat simulations similar
to ours but in the magnetohydrodynamical limit.
We have already set the stage for such a study because simulations
performed by Proga (2005) are a close  magnetohydrodynamical analog
of those presented here. We note that despite two different physical
regimes explored here and in our previous work, similarly
simple inflow/outflows 
solutions were obtained. One of the main motivation for 
this work, is to gain more insights to the general
problem of accretion and outflow production where many complex
time-dependent flow components are expected. It is also desirable
to connect numerical simulations with theoretical predictions such as  those
obtained by Blandford \& Begelman (1999) and 
Henriksen \& Valls-Gabaud (1994).

The SED and consequently gas temperature can be different from those 
we have in our simulations, even in an optically thin case. 
One of our simplifications is an assumption that the radiation temperature, 
$T_{\rm X}$ (or more generally the SED) does not change with $\theta$.
This assumption is motivated by X-ray observations showing that 
quasar radiation heats a low-density gas, 
on scales comparable to the Bondi radius,
up to an equilibrium Compton temperature of 
about $ 2\times10^7$~K (Sazonov et al. 2005; Allen et al. 2006
and references therein). However,
the SED and $T_{\rm X}$ can change with $\theta$. For example,
we assume that the X-ray/central object radiation does not change 
with $\theta$  (see eq. 7)
whereas the UV/disk radiation decreases with increasing $\theta$
(see eq. 6) .
Consequently one could expect the ratio between the X-ray and
UV flux to increase with increasing $\theta$. We plan in the near future, 
to test how important this effect  is and try to reconcile
the above theoretical expectation with observations.

We assume that a quasar is powered by accretion, but we do not couple 
the disk accretion rate, that determines radiation, to the mass accretion 
rate through the inner boundary. 
This $\MDOT_{\rm a}$--$\MDOT_{\rm in}(r_{\rm i})$ decoupling is physically 
motivated because an accretion disk is build from rotating gas and 
non-rotating gas will not significantly contribute to the disk formation
and disk radiation. In this sense we study radiatively inefficient flows
accreting onto an object with a radiatively efficient accretion disk.
Additionally, we note that
an accretion disk, which is located interior
to the inner radius of the computational domain, radiates
steadily during our simulations. The fact that the flow
in the computational domain settles into a steady state
within just a few dynamical time scales shows that 
our approach is reasonable.

Outflows in our simulations are collimated by the inflowing gas.
This collimation can lead to formation of a bottle-neck-like structure 
as we discovered in run B. There are two sources of concern about 
the collimation: is it an artifact of the outer boundary or of 
the numerical resolution.
We addressed this issue by rerunning simulation B and C
with higher resolution in the $\theta$ direction (100 grid points
instead 50). We found that the results  for the higher 
resolutions are very similar to those with the lower resolution
counterparts. The main difference was that the region
with $T<10^5$~K for $r> 1.8~\times~10^5~r_\ast$ is somewhat narrower (by
$\sim 1^\circ$) in the higher resolution run. 
We also found that the outer boundary are not responsible for the collimation
because our test calculations with  $r_{\rm o}$ four times larger
than in the runs presented in \S3 show collimation operating in a similar
way and far from the outer boundary. Additionally, we note
that in run C, collimation  begins at $r \sim 5\times 10^4~r_\ast$ 
(i.e., deep inside the computational domain) and is disconnected from
the outer boundary, and for that matter, from the inner boundary.
To study flow stability, in particular, stability of the outflow,
it is important to perform fully three-dimensional simulations
which will allow for non-axisymmetric effects and full development
of fluid instabilities.

We finish by returning to the questions we raised in 
\S 1. Our simulations show that AGN can have a substantial outflow which 
originates from the infalling gas. Such an outflow
can control the rate at which non-rotating matter is supplied
to the AGN central engine as its mass loss rate 
can be significantly higher than the mass inflow rate at small radii. 
For example, in run~C, as little as 10\% of the inflow at large radii  
reaches small radii because 90\% of the inflow is turned into an outflow. 
However, even the power of the strongest outflow 
is very low compared to the radiation power (i.e., for run C,
$P_{\rm k}/L=4\times 10^{-4}$). Finally, the gas 
in our simulations can be related to material
responsible for NELs, NALs, and warm absorbers. Additionally,
the inflowing gas could be the source of the gas that eventually
produces BALs and BELs at radii smaller than the inner radius
of our computational domain.
We plan to farther explore our model and relax  some of our assumptions
in order to confirm these conclusions.

ACKNOWLEDGMENTS:

We thank  G. Chartas, P. Hopkins, R. Kurosawa, T. Kallman,
J. Ostriker, and J. Stone for discussions.
We also thank an anonymous referee for comments that help us
clarify our presentaion.
This work is supported by NASA through grants 
HST-AR-10305 and HST-AR-10680 
from the Space Telescope Science Institute, 
which is operated by the Association of Universities for Research 
in Astronomy, Inc., under NASA contract NAS5-26555. Financial
support from DoE grant DE-FG52-06NA26217 is also acknowledged.

\clearpage

\begin{table*}
\scriptsize
\begin{center}
\caption{ Summary of results}
\begin{tabular}{l c c c c c c c c c c c   } \\ \hline
        & & & &  &    &      &      &                       &   &    \\
Run & $f_{\rm D}$ & $f_\ast$ & $f_{\rm UV}$ & $f_{\rm X} $ & $T_{\rm 0}$ &$\MDOT_{\rm in}(r_{\rm o})$ & $\MDOT_{\rm net}(r_{\rm i})$  & $\MDOT_{\rm out}(r_{\rm o})$   &     $v_r$         & $P_{\rm k}({\rm r}_o)$        &  $P_{\rm th}({\rm r}_o)$        \\ 
 & & & & & $2\times10^7$~K & $10^{25}$~g~s$^{-1}$      &  $10^{25}$~g~s$^{-1}$    &  $10^{25}$~g~s$^{-1}$   & $\rm km~s^{-1}$ &$10^{40}$~erg~s$^{-1}$ &  $10^{40}$~erg~s$^{-1}$   \\ \hline

A   &  0.5& 0.5  & 0.5& 0.5 &1     & -4   & - 1    & 3  & 700           & 2 & 4        \\

B   &  0.8& 0.2  & 0.8& 0.2 &1     & -8   &- 3     & 5  & 4000          & 100 & 0.8        \\

B1  &  0.8& 0.2  & 0.8& 0.2 &1/10  & -0.5 &- 0.09  &0.41& 1500          & 0.5 & 0.8        \\
B2  &  0.8& 0.2  & 0.8& 0.2 &1/3   & -2   &- 0.4   &1.6  & 1700         & 2 & 2        \\
B3  &  0.8& 0.2  & 0.8& 0.2 &3     & -9   &- 5     & 4  & 400           & 3 & 0.8        \\

C   &  0.95&0.05& 0.95&0.05 &1     & -9  & -1     & 8  & 6700           & 300 & 0.03    \\ \hline
\end{tabular}

\end{center}
\normalsize
\end{table*}

\newpage

\begin{figure}
\begin{picture}(280,540)
\put(0,0){\includegraphics{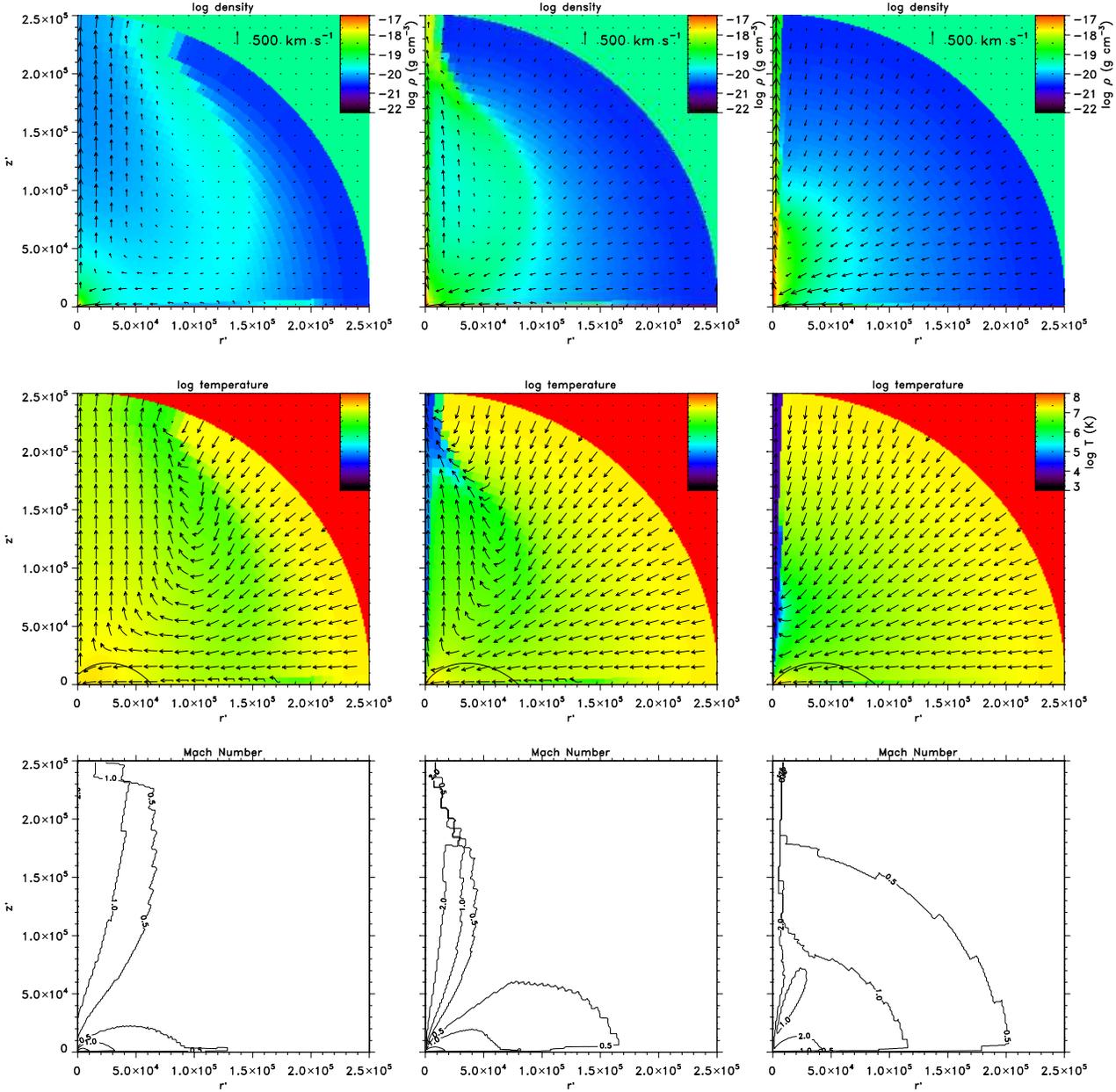}}
\end{picture}
\caption{ Comparison of the results for run A, B, and C (left, middle,
and right column, respectively).
{\it Top row of panels:}
Maps of logarithmic density overplotted by the 
poloidal velocity. For clarity, the arrows
are plotted with
the maximum velocity  set to 1000~${\rm km~s^{-1}}$.
{\it Middle row of panels:}
Maps of logarithmic temperature overplotted by the direction of the 
poloidal velocity. The solid curve in the bottom left corner
marks the position of the Compton radius corrected for the effects
of radiation pressure due to electron scattering (see eq. 19 in 
the main text).
{\it Bottom row of panels:}
Contours of the Mach number. 
The length scale is in units of the inner disk radius 
(i.e., $r' = r/r_\ast$ and $z' = z/r_\ast$).
}
\end{figure}

\begin{figure}
\begin{picture}(280,540)
\put(50,160){\includegraphics{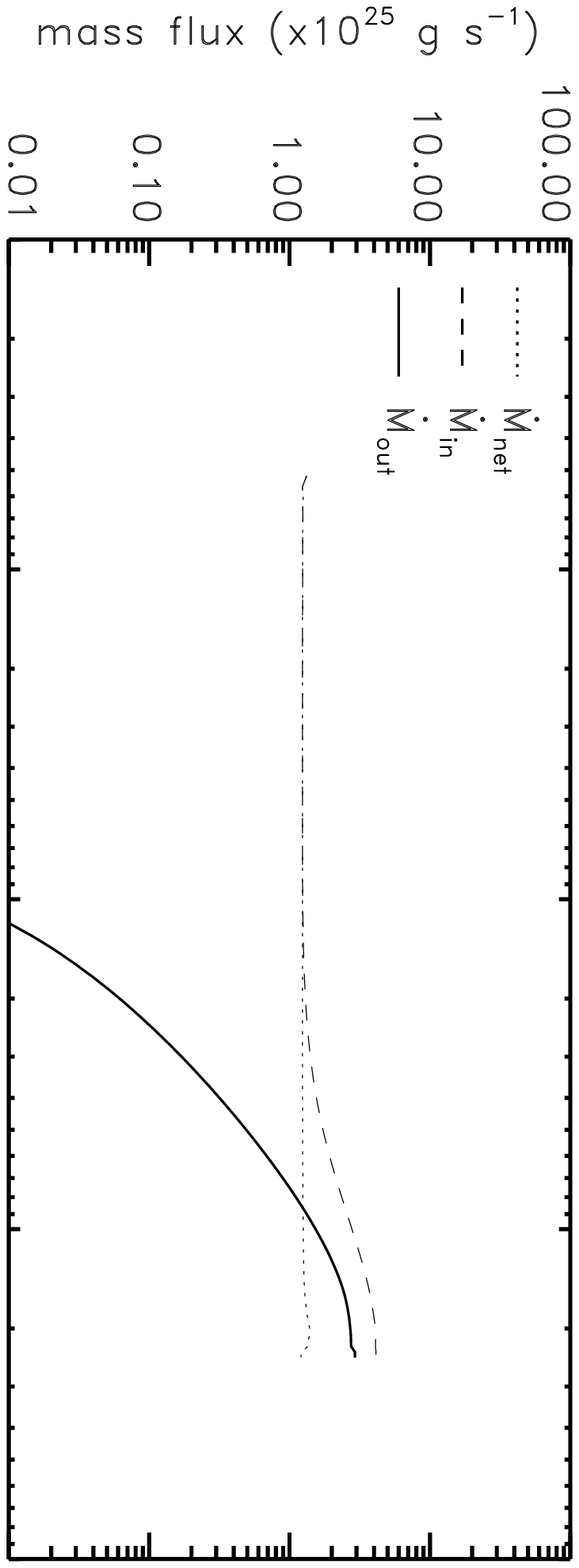}}

\put(50,0){\includegraphics{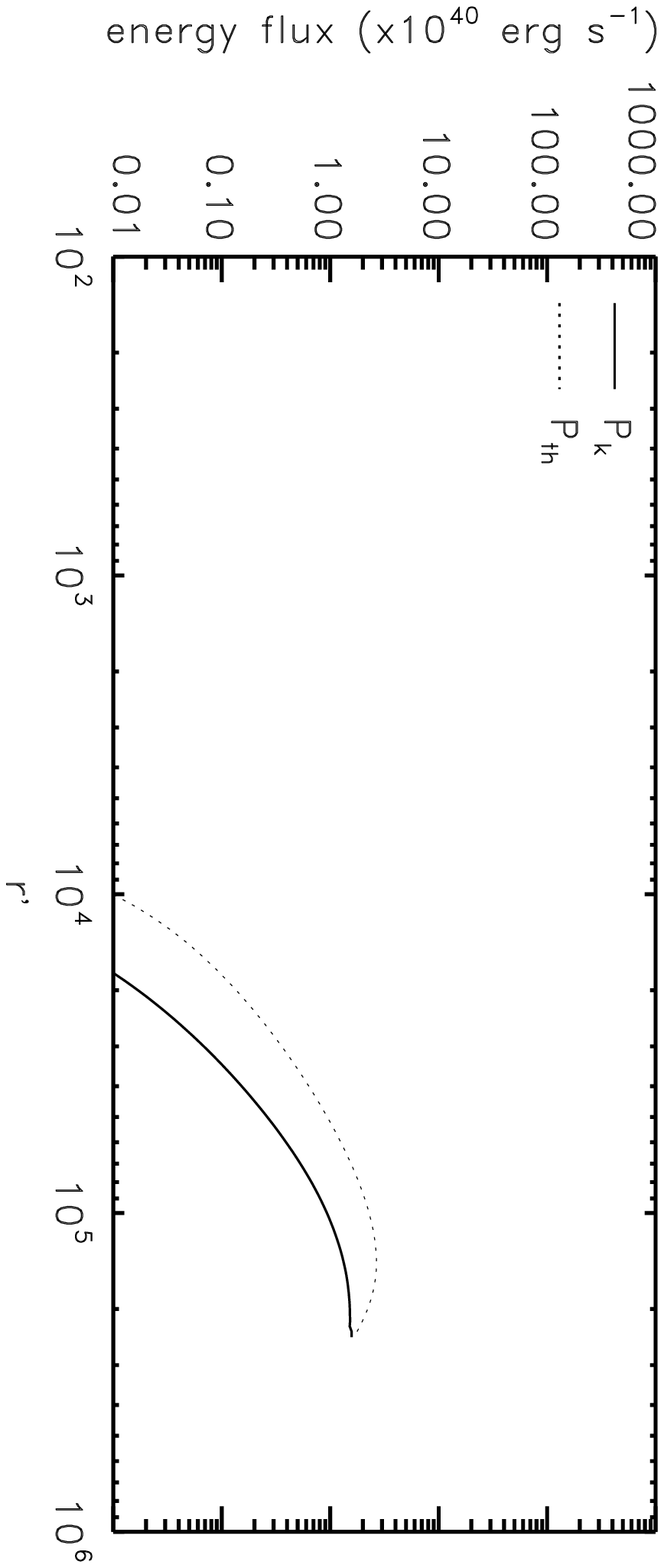}}
\end{picture}
\caption{{\it Top panel:} The mass flux rates 
as a function of radius for run A.
The solid, dashed, and dotted line corresponds to the outflow, inflow, 
and net rates, respectively (see eqs. 20, 21, and 22 for the formal
definitions). Note that the absolution value 
of the inflow and net rates are plotted because these quantities are negative.
{\it Bottom panel:} The energy fluxes carried out by the outflow
as a function of radius in run A1.
The solid and dashed line corresponds to the kinetic and thermal
energy flux, respectively (see eqs. 23 and 24 for the formal definitions).
The length scale is in units of the inner disk radius 
(i.e., $r' = r/r_\ast$).}
\end{figure}

\begin{figure}
\begin{picture}(280,590)
\put(50,200){\includegraphics{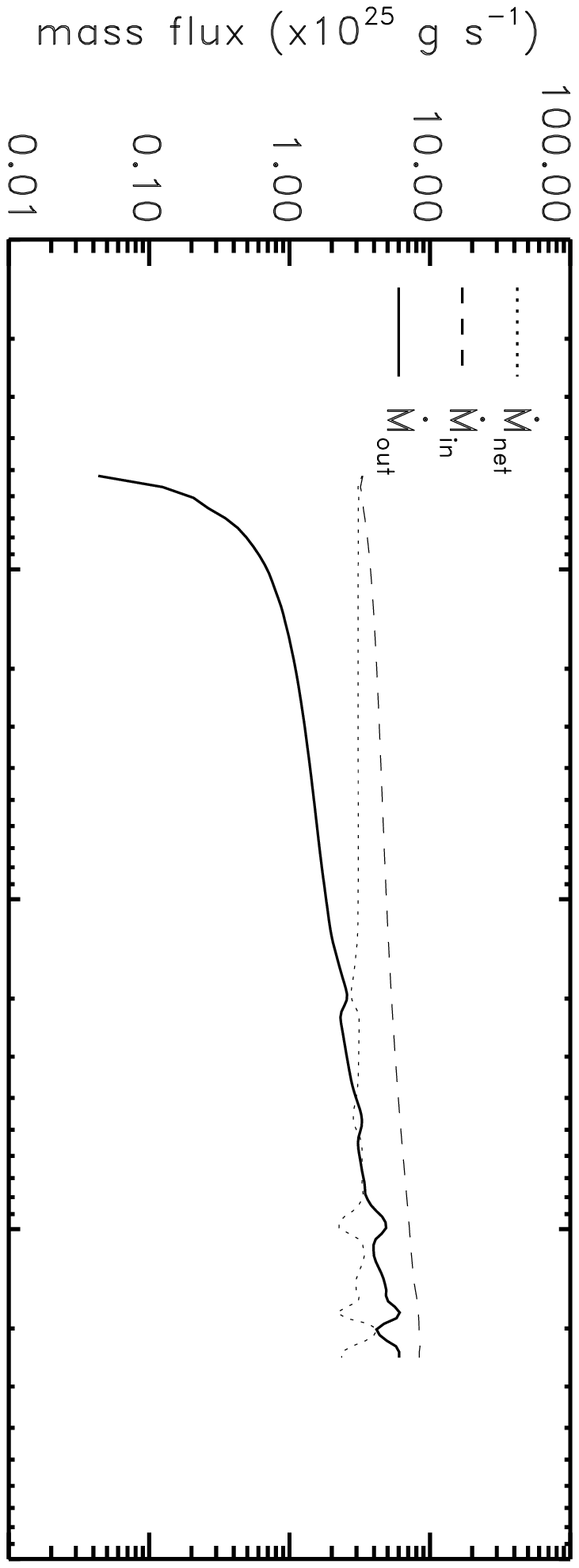}}

\put(50,40){\includegraphics{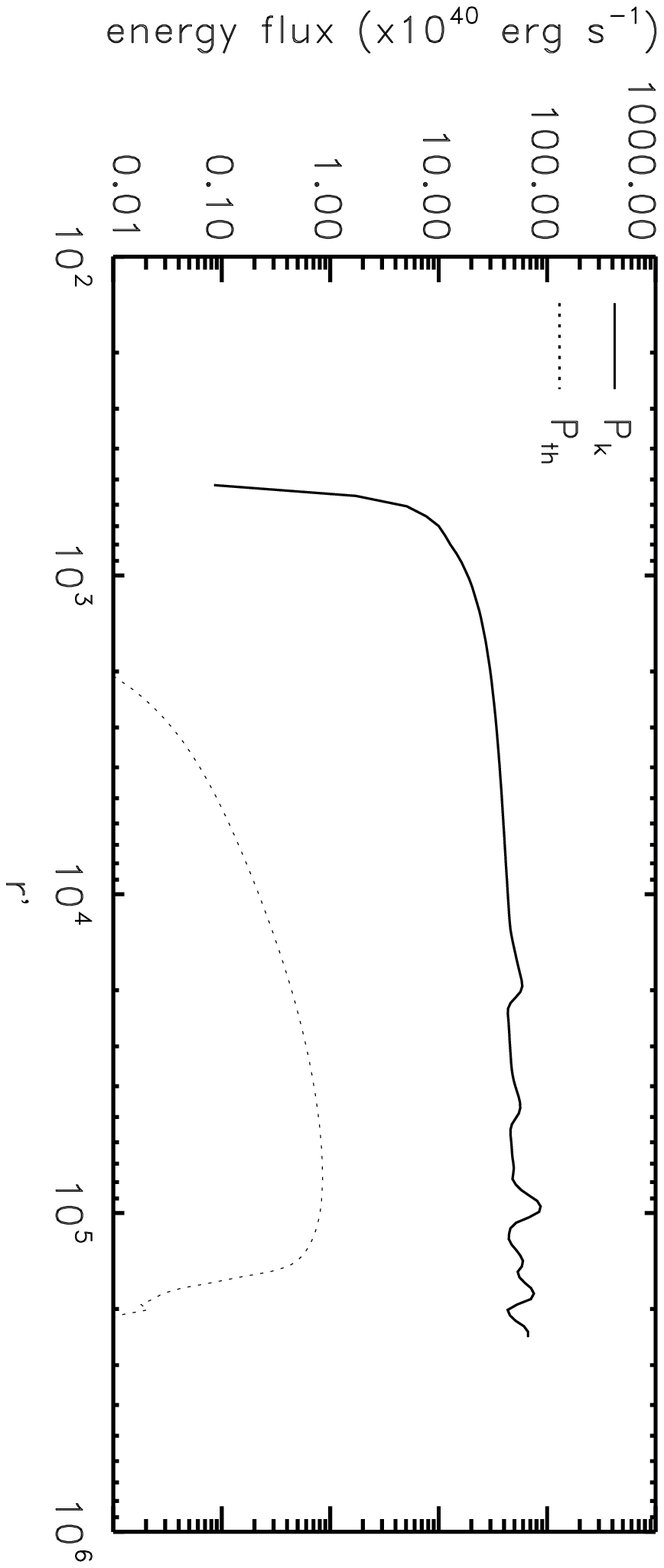}}
\end{picture}
\caption{ As Fig. 2 but for run B.}
\end{figure}

\begin{figure}
\begin{picture}(280,590)
\put(50,200){\includegraphics{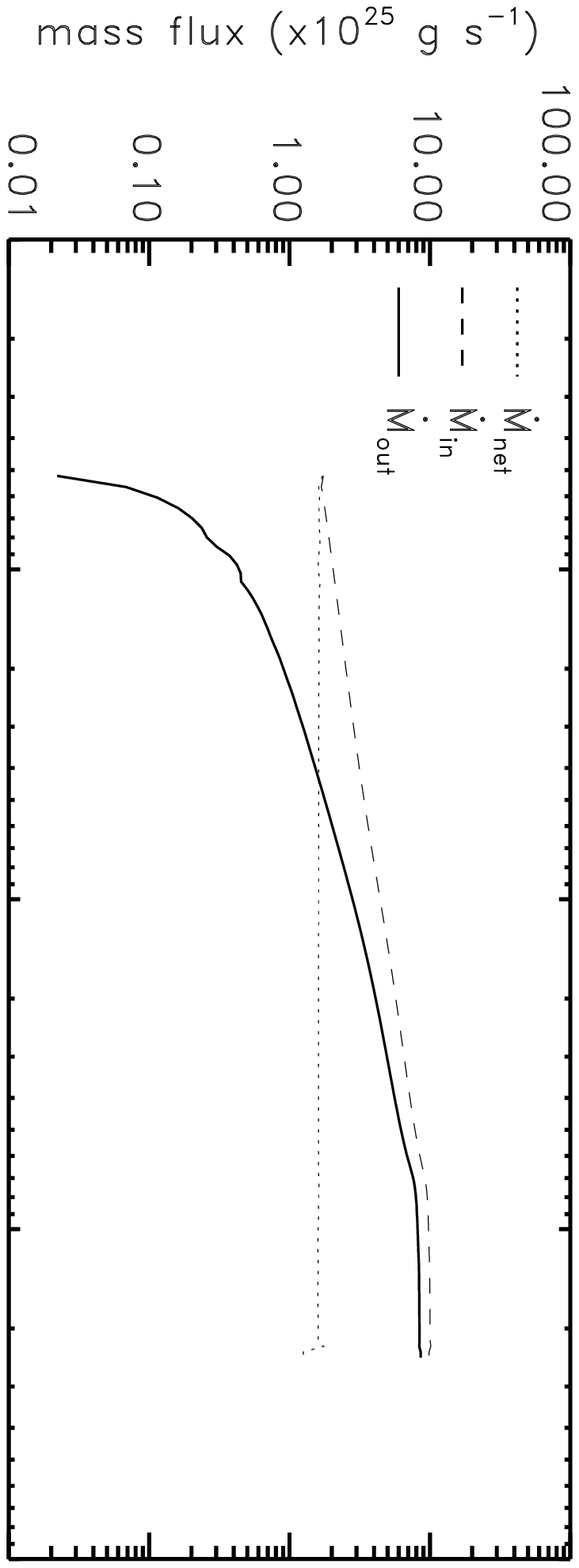}}

\put(50,40){\includegraphics{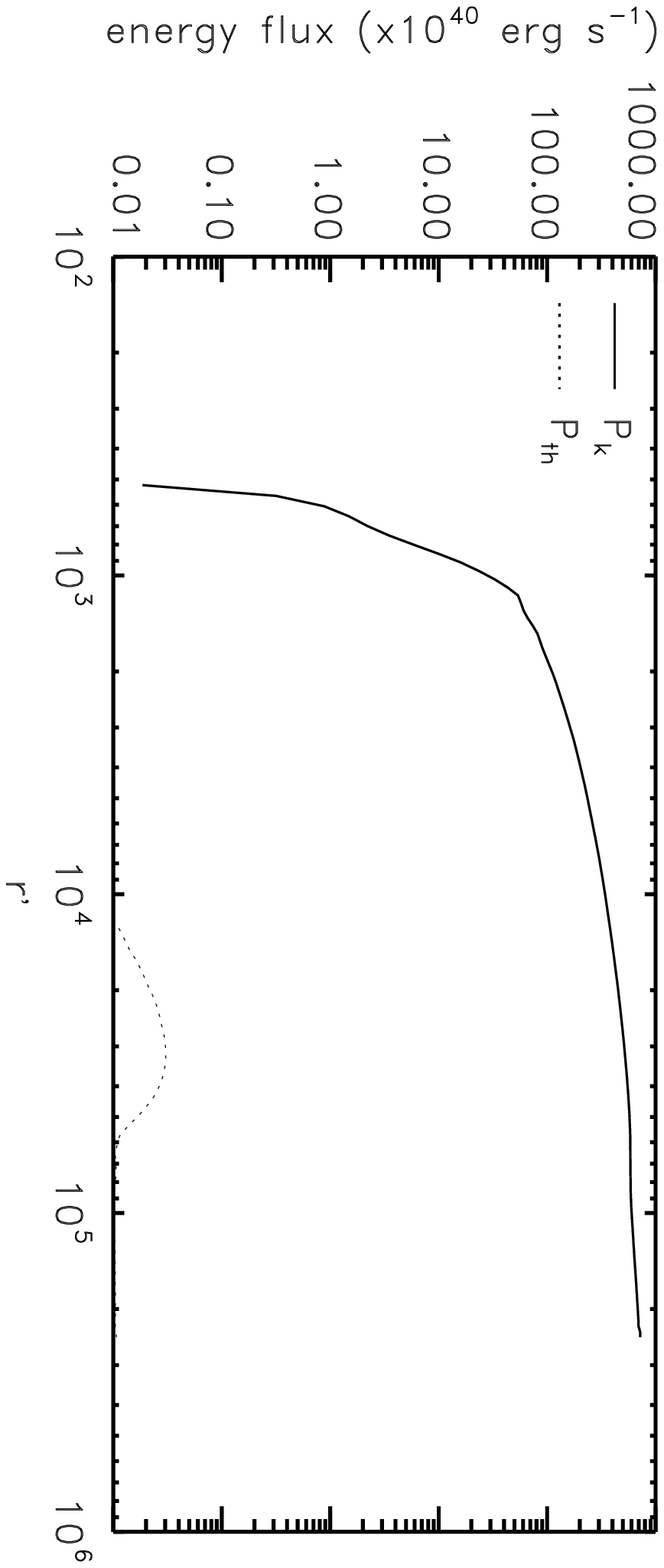}}
\end{picture}
\caption{ As Fig. 2 but for run C.}
\end{figure}

\end{document}